\documentclass[aps,showpacs,superscriptaddress,epsf,twocolumn]{revtex4-1}
\usepackage{graphicx}
\usepackage{dcolumn}
\usepackage[utf8]{inputenc}
\usepackage{tensor}
\usepackage{amsmath, amscd, amsthm, amssymb, mathrsfs,amsfonts}
\usepackage{amssymb}
\usepackage{bm}
\newcommand{\mybar}[1]{\setbox0\hbox{{#1}}
\makebox[\the\wd0][c]{%
\rule[0.42\ht0]{0.75\wd0}{0.7pt}}\hspace*{-\the\wd0}{#1}}
\usepackage{float}
\usepackage{hyperref}
\hypersetup{colorlinks=true, citecolor=blue, urlcolor=blue}
\usepackage{subfigure}
\usepackage{bm}
\usepackage{epstopdf}
\usepackage{amsthm}
\usepackage{float}
\graphicspath{{./figures/}}
\usepackage[toc]{appendix}
\usepackage{color,soul}
\usepackage[dvipsnames]{xcolor}
\usepackage[normalem]{ulem}

\def\R{\mathbb{R}}

\def\F{{\mathbf F}}

\def\e{{\mathbf e}}

\def\x{{\mathbf x}}
\def\y{{\mathbf y}}
\def\n{{\mathbf n}}
\def\F{{\mathbf F}}
\def\elle{\mathcal{L}}
\def\beq{\begin{equation}}
\def\eqn{\end{equation}\noindent}

\begin{document}

\title{Exactly solvable Stuart-Landau models in arbitrary dimensions}
\author{Pragjyotish Bhuyan Gogoi} 
\affiliation{Department of Physics and Astrophysics, University of Delhi, Delhi 110007, India}
\author{Rahul Ghosh} 
\affiliation{Department of Physics, Shiv Nadar IoE, Uttar Pradesh 201314, India}
\affiliation{Department of Chemistry, Indian Institute of Technology Delhi, Delhi 110016, India}
\author{Debashis Ghoshal} 
\affiliation{School of Physical Sciences, Jawaharlal Nehru University, Delhi 110067, India}
\author{Awadhesh Prasad }
\affiliation{Department of Physics and Astrophysics, University of Delhi, Delhi 110007, India}
\author{Ram Ramaswamy}
\affiliation{Department of Physics, Indian Institute of Technology Delhi, Delhi 110016, India}
\begin{abstract}
We use Clifford's geometric algebra to extend the  Stuart-Landau 
system to dimensions $D >2$ and give an exact solution of the 
oscillator equations in the general case. At the  supercritical 
Hopf bifurcation marked by a transition from stable fixed-point 
dynamics to oscillatory motion, the Jacobian matrix evaluated at 
the fixed point has $N=\lfloor{D/2}\rfloor$ pairs of complex 
conjugate eigenvalues which cross the imaginary axis simultaneously. 
For odd $D$ there is an additional purely real eigenvalue 
that does the same. Oscillatory dynamics is asymptotically confined 
to a hypersphere $\mathbb{S}^{D-1}$ and is characterised 
by extreme multistability, namely the coexistence of an infinite 
number of limiting orbits each of which has the geometry of a torus 
$\mathbb{T}^N$ on which the motion is either periodic or quasiperiodic. 
We also comment on similar Clifford extensions of other 
limit cycle oscillator systems and their generalisations.  
\end{abstract}
\maketitle

There is considerable current interest in higher-dimensional generalisations of 
oscillator models which provide realistic descriptions of natural systems 
\cite{Saber,Zhu2,Ott,Boccaletti2,Kumar}. Low dimensional linear and nonlinear 
oscillator models have traditionally been used to describe a variety of recurrent 
natural phenomena, and several dynamical systems that have periodic, aperiodic, or 
chaotic behaviour are well known \cite{Strog1,Strog2}.  
The normal form for Hopf bifurcation that provides a universal description of dynamics 
\cite{Stuart,Pant1,Poinsot} near the onset of oscillations is the Stuart-Landau system,
the equation of motion for which is given by
\beq
\label{eq:0}               
\dot{z}=(\varrho + i\omega -|z|^2)z,
\eqn 
where $z = x_1 + i x_2$ is a complex combination of the coordinates and $\varrho + i \omega$ 
is a complex control parameter that determines the nature of the dynamics \cite{fn1}. Closely 
related to the complex Ginsburg-Landau equation \cite{CGLE,Poinsot}, the Stuart-Landau system has
found wide application in studying collective phenomena in a wide variety of areas 
\cite{Pikovsky,Boccaletti,Pant1,Sune}. {\color{black}Of particular significance is the rotational symmetry of the model \cite{CGLE,Nirmal,AP,Wang}}.\\

In this Letter we discuss a consistent procedure for generalizing this system to arbitrary dimensions $D$ through the application of 
geometric (or Clifford) algebra \cite{HestenesSob,Doran}. 
The model being exactly solvable allows us to obtain a complete description of the dynamics. One major motivation is to study models 
of Hopf bifurcation, which has numerous applications \cite{Serrin,Marsden,Tyson,Hadeler,Wan}. 
{\color{black}In addition, the higher dimensional models have larger symmetry groups that play a crucial role in determining the nature of 
the dynamics}. The framework of Clifford algebra is a natural one in which these symmetries are apparent, moreover the 
role played by symmetries in $D>2$ can be analysed more completely by  breaking these partially and selectively \cite{fn2}.

\begin{center}
\begin{figure}[htp]
     \includegraphics[width=0.38\textwidth]{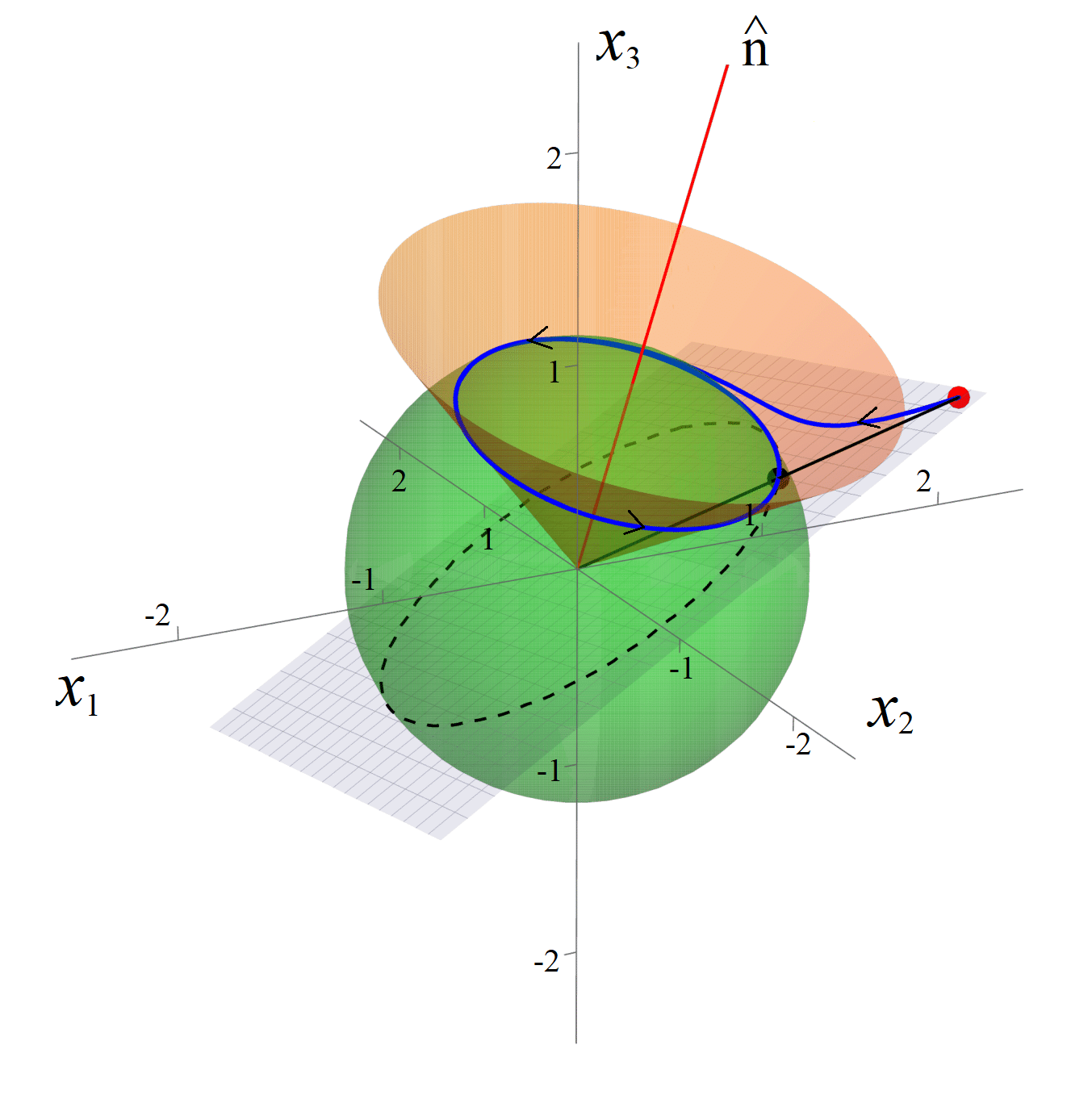}
     \caption{(Color online) The $D=3$ Stuart-Landau system with parameters $\big(\mu_{12}, \mu_{23}, \mu_{13}\big) = 
     (-0.3,-0.2,0.1)$. For any initial $\x(0)$ (marked by a red dot) on the cone the system described by Eq.~\eqref{eq:12} asymptotes to the circle 
     $\mathbb{S}^1$ (solid blue line) at which the cone intersects the sphere $\mathbb{S}^2$. The normal mode coordinate $y_3$ (red) 
     is along $\hat{\n}$, the axis of the cone, while $y_1, y_2$ span the plane perpendicular to it. 
    The initial bivector-plane  $\mathsf{P}_2(0)$ (shown in blue) intersects $\mathbb{S}^2$ in a great circle, $\mathcal{C}(0)$ (dashed black line). The ray from the origin to $\x(0)$ 
    intersects $\mathcal{C}(0)$ and $\mathbb{S}^1$ at a common point marked as a black dot.}
     \label{fig:1}
\end{figure}
\end{center}

In Clifford algebra, an orthonormal set of unit vectors $\{\e_i\}$ in a $D$-dimensional Euclidean vector space $\mathbb{R}^D$ satisfy  
\cite{HestenesSob,Doran}
\beq 
\{\e_{i},\e_{j}\} = \e_i\e_j + \e_j\e_i = 2\delta_{ij}, \: i,j = 1,\cdots, D.
\label{eq:CliffordRelns}
\eqn 
The geometric product of the basis vectors, defined as 
\beq
\e_i \e_j = \delta_{ij} + \frac{1}{2}(\e_i\e_j - \e_j\e_i) \equiv \delta_{ij} + \e_i\wedge\e_j,
\label{eq:GeomProd}
\eqn
is the sum of a scalar and a bivector, the latter arising from the antisymmetric part of 
the product. The Clifford algebra is 
generated by the free product of the vectors, modulo the Clifford relations 
Eq.~\eqref{eq:CliffordRelns}.  Since 
$(\e_i\e_j)^2=-1$ for $i\ne j$, the ${D \choose 2}$ basis bivectors behave like imaginary units.
Indeed, this is the extension of the imaginary unit, of which Hamilton's quaternions are a special case.\\

{\color{black}The Stuart-Landau oscillator \eqref{eq:0} can clearly be generalised to any higher dimension $D$ in many different ways. In order to 
identify the properties that would be desirable to retain in a generalisation, we first rewrite Eq.~(\ref{eq:0}) in the vector form,
\beq
\Dot{\x}=(\varrho-|\x|^2)\x + \mathbb{A}\x,
\label{eq:100}
\eqn
where $\x = x_1\e_1 + x_2\e_2 = (x_1 \; x_2)^T$ is the coordinate vector and 
$\mathbb{A}=\left(\begin{array}{cc}
0 & -\omega\\
\omega & 0
\end{array}\right)$ 
is an anti-symmetric matrix. 
The first term $(\varrho-|x|^2)\x$ points along $\x$ and involves dissipation. The second term $\mathbb{A}\x$ is 
perpendicular to $\x$, since $(\mathbb{A}\x)\cdot\x=0$. Consequently, the magnitude $|\x|$ evolves independently to assume its asymptotic 
value leading to a limit cycle $|\x|=\sqrt{\varrho}$ along which $\mathbb{A}\x$ generates rotational motion. The system has two-
dimensional rotational symmetry: $z\rightarrow e^\alpha z$ in Eq.\eqref{eq:0} or $\x\rightarrow \x\mathbb{R}$, where 
$\mathbb{R}\in\text{SO}(2)$ in Eq.\eqref{eq:100}. Use of Clifford algebra facilitates preservation of both these essential aspects in arbitrary dimensions.\\ 

The most general form of the linear part in \eqref{eq:100} is $\sum_{jk} \mathbb{T}_{jk}x_k$, where $\mathbb{T}$ is a 
$D\times D$ matrix, which we may further express as a sum of a symmetric matrix $\mathbb{S} = \tfrac{1}{2}(\mathbb{T}+\mathbb{T}^T)$ 
and an antisymmetric matrix $\mathbb{A} = \tfrac{1}{2}(\mathbb{T} - \mathbb{T}^T)$, i.e., $\mathbb{T}_{jk} = \mathbb{S}_{jk} + \mathbb{A}_{jk}$. 
By use of the Clifford relations \eqref{eq:CliffordRelns}, we set $\mathbb{S}_{jk}=\frac{1}{D}(\text{Tr }\mathbb{T})\delta_{jk}$, where $\frac{1}{D}\text{Tr }\mathbb{T} = \varrho$ is a real 
scalar, and the components of $\mathbb{A}$ are identified with the components of a bivector. This leads to a generalised Stuart-Landau equation in $D$ dimension
\begin{equation}
\Dot{\x} = (\varrho - |\x|^2) \x + \x \cdot \boldsymbol{\mu},
\label{eq:2}
\end{equation}
where $\x = \displaystyle{\sum_{i=1}^{D}} x_i\e_i$ is the coordinate vector, $\boldsymbol{\mu} = \displaystyle{\sum_{i < j}} \mu_{ij} \e_j\e_i$
is a bivector, the components $\mu_{ij}$ of which are (real) parameters. Note that the dot product
$\x\cdot\boldsymbol{\mu} = \frac{1}{2} \left(\x\boldsymbol{\mu} - \boldsymbol{\mu}\x\right)$ is a vector is orthogonal to $\x$, since 
$(\x\cdot\boldsymbol{\mu})\cdot\x=0$, it leads to rotations in Eq.\eqref{eq:2}. Importantly, this also makes the equation separable and 
consequently exactly solvable, as we explore later in detail. The form of  the equation remains the same when it is right-multiplied by a 
$D\times D$ orthogonal matrix corresponding to rotation and reflection, which is manifested by the conjugate action 
of products of vectors in Clifford algebra. In other words, the equation is covariant under a right action of the group O($D$), thus 
preserving the symmetries of Eq.\eqref{eq:100}. Equivalently, Eq.\eqref{eq:2} in components lead to 
\begin{equation}
\dot{x}_i = (\varrho - r^2) x_i + \sum_{i,j} \mathbb{M}_{ij} x_j,
\label{eq:4}
\end{equation}
where $r^2=|\x|^2$, and the association of the anti-symmetric matrix $\mathbb{M}$ with off-diagonal elements $\mathbb{M}_{ij} = \mu_{ij} = -\mathbb{M}_{ji}$ is thanks to the identity 
\begin{equation*}
\x \cdot \boldsymbol{\mu} = \sum_{\substack{i,j=1 \\ i<j}}^{D} \mu_{ij} (x_j\e_i-x_i\e_j) = \sum_{i,j} \mathbb{M}_{ij} x_j\e_i,
\end{equation*}
It is easy to verify that for $D=2$, we recover the real and imaginary parts of Eq.~\eqref{eq:0}, and similarly Eq. \eqref{eq:100}, with $\mu_{ij}=\omega$.}\\ 

At the fixed point $\x = 0$, the Jacobian $J$ for Eq.~\eqref{eq:2} consists of the 
elements $J_{ij} = \varrho\,\delta_{ij} + \mu_{ij}$. 
All $D$ eigenvalues have the same real part $\varrho$, while imaginary parts are 
independent of $\varrho$. As $\varrho$ is 
varied from negative to positive values, eigenvalues cross the imaginary axis 
symmetrically, destabilising the fixed point at the 
origin and leading to oscillatory motion: this characterises Hopf bifurcation \cite{Marsden}. 
Additionally, these eigenvalues come in 
$N = \lfloor{D/2}\rfloor$ {\em complex conjugate} pairs when $D$ is even, 
while for $D$ odd there is an additional {\em purely 
real} eigenvalue. This leads to additional fixed points and differences in the 
geometry of the attractors in odd dimensions.\\

The  model in Eq.~\eqref{eq:2} retains the property of being exactly solvable.
Since $\mathbb{M}$ is antisymmetric, an orthogonal  transformation 
$\R \in \mathrm{SO}(D)$ can be found to bring it to the Jordan canonical form: $\R\mathbb{M}\R^{\mathsf{T}} = 
\mathbb{M}_{\mathrm{J}}$, a block-diagonal matrix composed of $2\times 2$ blocks of the form
$\left(\begin{array}{cc}
0 & -\omega_j\\
\omega_j & 0
\end{array}\right)$,
$j=1,\cdots, N$. 
For odd $D$, there is an additional row and column of zeroes corresponding to the additional 
zero-frequency normal mode, for 
which there are two additional fixed points, namely the points of intersection of the 
corresponding eigendirection with the 
hypersphere $\mathbb{S}^{D-1}$. In terms of the transformed vector $\y = \R\, \x$, Eq.~\eqref{eq:4} becomes
\begin{equation}
\dot{y}_i = \sum_j \left( (\varrho - r^2) \delta_{ij} + (M_{\mathrm{J}})_{ij}\right) y_j.
\label{eq:yEOM}
\end{equation}
since $r^2 = |\x|^2 = |\y|^2$ is an invariant of SO($D$). Due to the block-diagonal form of 
the matrix, the pairs of coordinates 
$(y_{2j-1}, y_{2j}), j=1,\cdots,N$ (plus the additional unpaired coordinate $y_D$ for odd $D$) 
evolve independently. For each pair, at any time we have 
\begin{equation*}
\left(\begin{array}{c}
\dot{y}_{2j-1}\\
\dot{y}_{2j}
\end{array}\right) =
\left(\begin{array}{cc}
\varrho - r^2 & -\omega_j\\
\omega_j & \varrho - r^2 
\end{array}\right)
\left(\begin{array}{c}
{y}_{2j-1}\\
{y}_{2j}
\end{array}\right),
\end{equation*}
or more compactly in the complex coordinates $z_j = y_{2j-1} + i y_{2j}$,
\begin{equation}
\dot{z}_j 
= \left( (\varrho - r^2(t)) + i \omega_j \right) z_j.
\label{eq:ComplexEOM}
\end{equation}
{\color{black}The coordinates $z_j$ are clearly separable, and therefore the modes evolve independently and do not interact.} 
Using the identity $(\x\cdot\boldsymbol{\mu})\cdot\x=0$, we now solve for $r$ and obtain
\begin{equation*}
r(t) = \sqrt{\varrho} \left(1 - C e^{-2\varrho t}\right)^{-1/2}
\end{equation*}
where $C = (1 - \varrho/r_0^2)$ is a constant determined by the initial condition $r_0 = r(0)$. Notice that even in $D$ dimensions, 
$r$ asymptotes to the limiting value $r_*=\sqrt{\varrho}$, with a characteristic time $\mathcal{O}(1/(2\varrho))$, as it does in $D=2$. 
Upon integration Eq.\eqref{eq:ComplexEOM} leads to
\begin{equation}
z_j(t) = 
\mathsf{a}_j^{(\infty)}\, e^{i\omega_j t}
\left(1 - C e^{-2\varrho t}\right)^{-1/2}
\label{eq:Radj}
\end{equation}
and for odd $D$, $y_D(t) = \mathsf{a}_D^{(\infty)}\left(1 - C e^{-2\varrho t}\right)^{-1/2}$ for the unpaired (real) coordinate. 
Thus, we see that as the system approaches the hypersphere $r = \sqrt{\varrho}$ asymptotically, the trajectory on the $z_j$ plane 
approaches a circle $\mathbb{S}^1_j$ of radius $\mathsf{a}_j^{(\infty)}$; this is also reminiscent of the constant energy 
trajectory in the phase space of a simple harmonic oscillator. These $N$ quantities, namely $\mathsf{a}_j^{(\infty)}$, $j=1,\ldots, N$, and 
for odd $D$, an additional one, $\mathsf{a}_D^{(\infty)}$, therefore constitute asymptotic conserved quantities, and are determined 
from the initial conditions. The constants $\mathsf{a}_j^{(\infty)}$ are furthermore constrained by the condition 
$\sum_j \big(\mathsf{a}_j^{(\infty)}\big)^2 = \varrho$, where the sum also includes the unpaired real coordinate when appropriate.\\

At any instant, the position and the velocity defines the bivector-plane $\mathsf{P}_2(t)$ through the origin, corresponding to the 
bivector $\x\wedge\dot{\x}$. The time evolution of the unit bivector $\boldsymbol{\elle} \equiv (\x\wedge\dot{\x})/|\x\wedge\dot{\x}|$ 
corresponding to this plane can be shown to be
\begin{equation} 
\frac{d}{dt}\boldsymbol{\elle} = \frac{1}{|\x|\;|\x.\boldsymbol{\mu}|} \sum_{i,j,k}x_i(\mathbb{M}^2)_{jk} x_k\,\e_i\wedge\e_j.
\label{eq:bplane}
\end{equation}
The initial bivector-plane $\mathsf{P}_2(0)$ intersects the limiting sphere $\mathbb{S}^{D-1}$ along a great circle $\mathcal{C}(0)$ (cf.~Fig~\ref{fig:1}). 
In order to quantify the asymptotic trajectory, it is convenient to introduce the vectors $\boldsymbol{\xi}(t) = \sqrt{\varrho}\,\x(t)/r(t)$ and $\boldsymbol{\eta}(t) = \sqrt{\varrho}\,\y(t)/r(t)$.
The system in Eq.\eqref{eq:2}, like the original Stuart-Landau oscillator \eqref{eq:0}, is asymptotically linear. Indeed, one can 
easily verify that $\dot{\boldsymbol{\xi}} = \boldsymbol{\xi\cdot\mu}$ is the same as the asymptotic form of Eq.~\eqref{eq:2}. Likewise,
$\dot{\boldsymbol{\eta}} = \boldsymbol{\eta}\cdot\boldsymbol{\mu}_{\text{J}}$.
The line joining the origin to the initial point $\y(0)$ intersects $\mathbb{S}^{D-1}$ at $\boldsymbol{\eta}_0 = \boldsymbol\eta(0)$. Thus 
this point is on the great circle, as well as on the final asymptotic trajectory (cf.~Fig~\ref{fig:1}). The circle $\mathbb{S}^1_j$ is in fact inscribed in the projection 
of the great circle $\mathcal{C}(t)$ to the $z_j$-plane. Defining  $\zeta_j=\sqrt{\varrho}\, z_j/r$, $j=1,\cdots,N$, one obtains the values of the asymptotically conserved quantities as
\begin{eqnarray}
\mathsf{a}_j^{(\infty)} &=& |\zeta_j(0)| 
,\: j=1,\cdots,N\nonumber\\ 
\big(\mathsf{a}_D^{(\infty)} &=& |\eta_D(0)|~~ \mathrm{for~odd}~D\big )   
\label{eq:Conserved}
\end{eqnarray}
and these are clearly related to the initial conditions.\\

{\color{black}Each initial point leads to a torus attractor on which there are, asymptotically,
a set of $N$ conserved quantities of the system. These are the radii $(\mathsf{a}_1^{(\infty)}, 
\cdots, \mathsf{a}_N^{(\infty)})$ of the $N$ independent 
cycles on the  torus $\mathbb{T}^{N} = \mathbb{S}^1_1\times\cdots\times\mathbb{S}^1_N$. The existence of an infinite set of attractors is the defining characteristic
of systems with {\em extremely multistability} \cite{Showalter}}. The basin of attraction for each
attractor is a generalised cone that is generated by the rays joining the origin to the points on $\mathbb{T}^{N}$ which is its base \cite{fn3}. For 
$D$ odd, the axis of symmetry of the cone is along $y_D$, the normal mode corresponding to the zero eigenvalue of $\mathbb{M}$. 
{\color{black}It is worth emphasizing that the limiting orbits are not isolated in this case. However, since for each one of there is a basin 
of attraction, we shall continue to refer to them as limit cycles, while noting their non-isolated nature.} \\

Since the motion arises out of the dynamics of the individual modes, the detailed nature 
of a trajectory will depend on the ratios 
of the $\omega$'s, as well as on the initial conditions. For $\varrho>0$, the origin 
is an unstable fixed point for all $D$. In $D=1$, 
Eq.~\eqref{eq:2} has two stable fixed points $x_{1*} = \pm \sqrt{\varrho}$, namely a 
zero-dimensional sphere $\mathbb{S}^0$. In the Stuart-Landau system \cite{Stuart} in $D=2$, 
Eq.~\eqref{eq:2} specifies a limit cycle $\mathbb{S}^1$ defined 
by $|\x| = |\y| = \sqrt{\varrho}$, which is traversed with an angular frequency $\omega$.\\ 

In $D=3$, there are a conjugate pair of angular frequencies $\pm\omega$ (as in $D=2$) 
and an additional zero eigenvalue of
$\mathbb{M}$. In three dimensions there is a unique normal vector for an oriented plane,
so a bivector can be seen as being dual to a vector. Specifically,
$\boldsymbol{\mu}  = \mu_{12}\e_2\e_1 + \mu_{23} \e_3\e_2 + \mu_{13} \e_3\e_1$
will be dual to 
$\n= -\mu_{23} \e_1 + \mu_{13} \e_2 - \mu_{12}\e_3$.
Hence the system evolves according to
\beq  
\Dot{\x}=(\varrho-|\x|^2)\x +\n\times\x, 
\label{eq:12}
\eqn 
which in terms of the asymptotic coordinate simplifies to $\dot{\boldsymbol{\xi}} = \n\times \boldsymbol{\xi}$. Thus 
asymptotically the motion is counterclockwise on a circle $\mathbb{S}^1$ that lies on the invariant manifold 
$\mathbb{S}^2$ defined by $|\x|^2 = \varrho$ \cite{Ott} around the vector $\n$ with an angular frequency $|\n| = \omega = \sqrt{ \mu_{12}^2 + \mu_{23}^2 + \mu_{13}^2}$, 
as shown in Fig.~\ref{fig:1}. The limiting orbit $\mathbb{S}^1$ of radius $\sqrt{\varrho(1- (\boldsymbol{\hat{\x}}_0\cdot\hat{\n})^2)}$ 
lies on the plane $\boldsymbol{\xi} \cdot \n = \text{constant}$. {\color{black}The associated vector $\n$ makes it possible for these three-dimensional 
oscillators to be utilised as oscillating, orientable individual agents in swarm dynamics \cite{Ott,Kumar}}. \\

The basin of attraction of the limiting orbit $\mathbb{S}^1$ is the cone defined by the rays joining the origin to the points of 
$\mathbb{S}^1$ (see Fig.~\ref{fig:1}). The orbit is stable to perturbations along the surface of the cone, but perturbations in 
transverse directions lead to other orbits. The opening angle of the cone, $\cos^{-1}({\hat{\x}_0\cdot\hat{\n}})$ is determined by the 
initial point: there is thus a one-parameter family of limit cycles that can be labelled by this angle, leading to the extreme 
multistability (See Supplementary Material \S{1}). There are two additional fixed points at $\pm\sqrt{\varrho}(n_1, n_2, n_3)/|\mathbf{n}|$, 
namely the points where the ray $\n$ intersects the invariant sphere (this corresponds to the degenerate cone of vanishing angle) which are 
stable for perturbations along $\n$. Transitions to multistable periodic states are usually linked with a degenerate Hopf bifurcation
when the transversality criterion is not satisfied at the bifurcation \cite{Shiau,Meer}. However, transversality is maintained here. The 
eigenvector corresponding to the zero eigenvalue of $\mathbb{M}$ is $\n$ itself, while the other two eigenvectors corresponding to the 
eigenvalues $\pm i\omega$ span the plane perpendicular to $\n$. \\

The $D=2$ and $D=3$ cases are prototypical of the even and odd $D$ respectively. 
Further, the dynamics of $D=(2N+1)$ case is quite closely related to that of $D=2N$. For odd $D$ there are two new fixed points corresponding to the additional
zero eigenvalue of $\mathbb{M}$, which are attractive along the associated eigendirection. This pair (which are at the intersection of the 
corresponding eigenvector with $\mathbb{S}^{D-1}$) together with the origin is similar to
the $D=1$ system. In both cases, the asymptotic trajectory lies on an 
$N$-dimensional torus, $\mathbb{T}^{N} = (\mathbb{S}^1)^{N} \subset \mathbb{S}^{D-1}$, and the basin of attraction is 
a generalised cone, as described above. For even $D$ there is no preferred direction 
for the axis of symmetry of the cone, but in odd dimensions, 
the axis of symmetry is along the eigenvector corresponding to the zero eigenvalue, 
as seen in the $D=3$ case. \\

\begin{figure}[htp]
      \centering
     \includegraphics[width=0.36\textwidth]{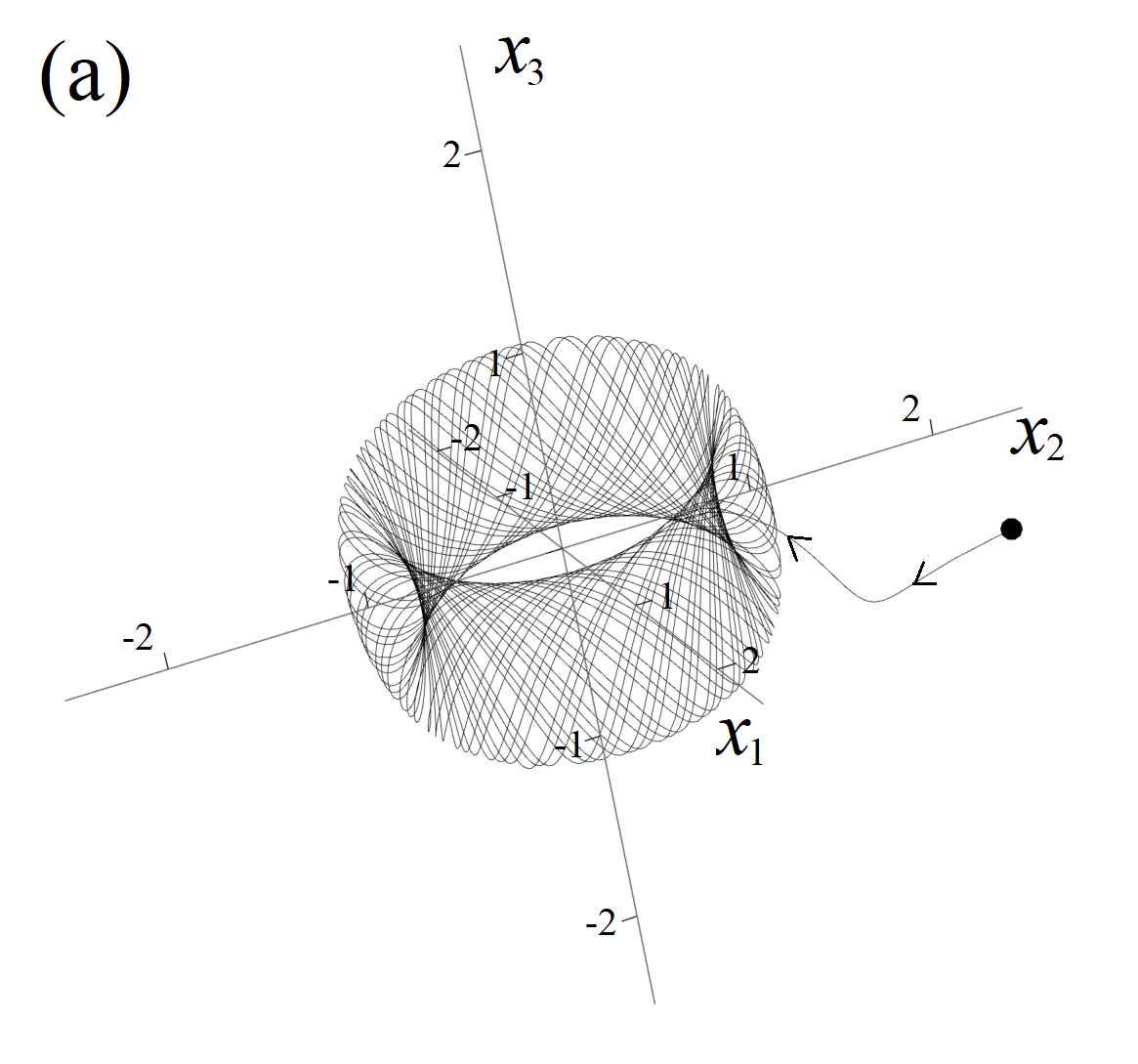}
     \includegraphics[width=0.33\textwidth]{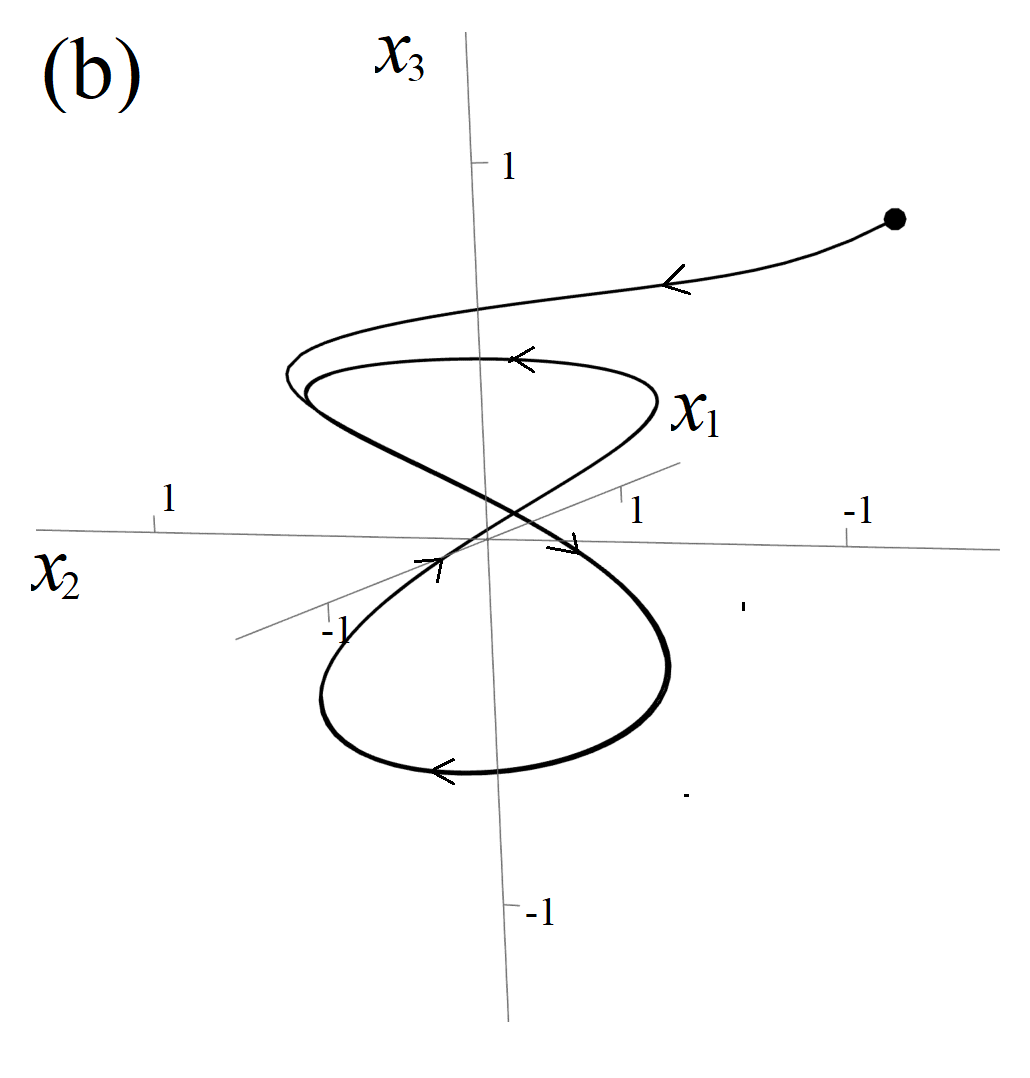}
     \caption{Trajectories in the $D=4$ Stuart-Landau system for typical values of $\mu_{ij}$ are
     quasiperiodic; a projection is shown in (a). If the ratio of mode frequencies
     is rational then the orbit is a closed loop as shown in (b) when $\omega_1/\omega_2$ = 2.}
      \label{fig:2}
\end{figure}

The detailed nature of the trajectories depend on the ratios of the angular 
frequencies determined by $\mu_{ij}$. In the generic case the ratios 
will be irrational leading to quasiperiodic orbits that ergodically cover the 
torus attractor $\mathbb{T}^{N}$. The three-dimensional projections of such a 
quasiperiodic trajectory for $D=4$ are shown in Fig.~\ref{fig:2}(a).
If the ratios of the angular frequencies are rational, the trajectory on $\mathbb{T}^{N}$ 
closes on itself after a finite number of rotations (see Fig.~\ref{fig:2}(b)).\\

The spectrum of Lyapunov exponents changes at the bifurcation from being all negative for 
$\varrho < 0$ to being a single negative 
exponent and $D-1$ zero exponents for $\varrho > 0$. In particular, for $D=3$, the Lyapunov 
spectrum has the signature 
$(0,0,-)$ after Hopf bifurcation. Although usually associated with quasiperiodicity 
\cite{Fiedler,Kuznetsov}, here it corresponds 
to a periodic orbit. The zero components are associated with the directions 
tangential to the invariant sphere which is foliated 
by the one-parameter family of limit cycles while the negative exponent 
corresponds to the direction of stability, namely the normal. \\

An additional interesting possibility for a consistent truncation arises in 
$D=4$. In the generic case the motion is asymptotically confined to a 
torus $\mathbb{T}^2 \subset \mathbb{S}^3$, 
and the system has O(4) symmetry, namely the group of rotations and reflections in $D=4$. 
But since O(4) $\sim$ O(3)
$\times$O(3), we can identify the three quantities 
\begin{equation*}
\mu_{i4} = \pm \frac{1}{2} \sum_{jk} \epsilon_{ijk}\mu_{jk} \equiv \nu_i,\: i=1,2,3,
\end{equation*}
to reduce the number of independent parameters from six to three, effectively restricting to 
the {\em even} subalgebra of the Clifford algebra: these are Hamilton's quaternions that 
describe rotations in {\em three} dimensions. Note that the diagonal SO(3)$\sim$ U(2) symmetry 
is explicit in terms of the complex coordinates, e.g., $x_1 + i x_4$ and $x_2 - i x_3$. 
Similar consistent truncation may be possible in other dimensions as well. In $D=7$
for example, one may choose relations between $\mu_{ij}$ to restrict the symmetry 
group to $G_2 \subset \text{SO}(7)$.
\begin{center}
\begin{figure}[htp]
     \includegraphics[width=0.38\textwidth]{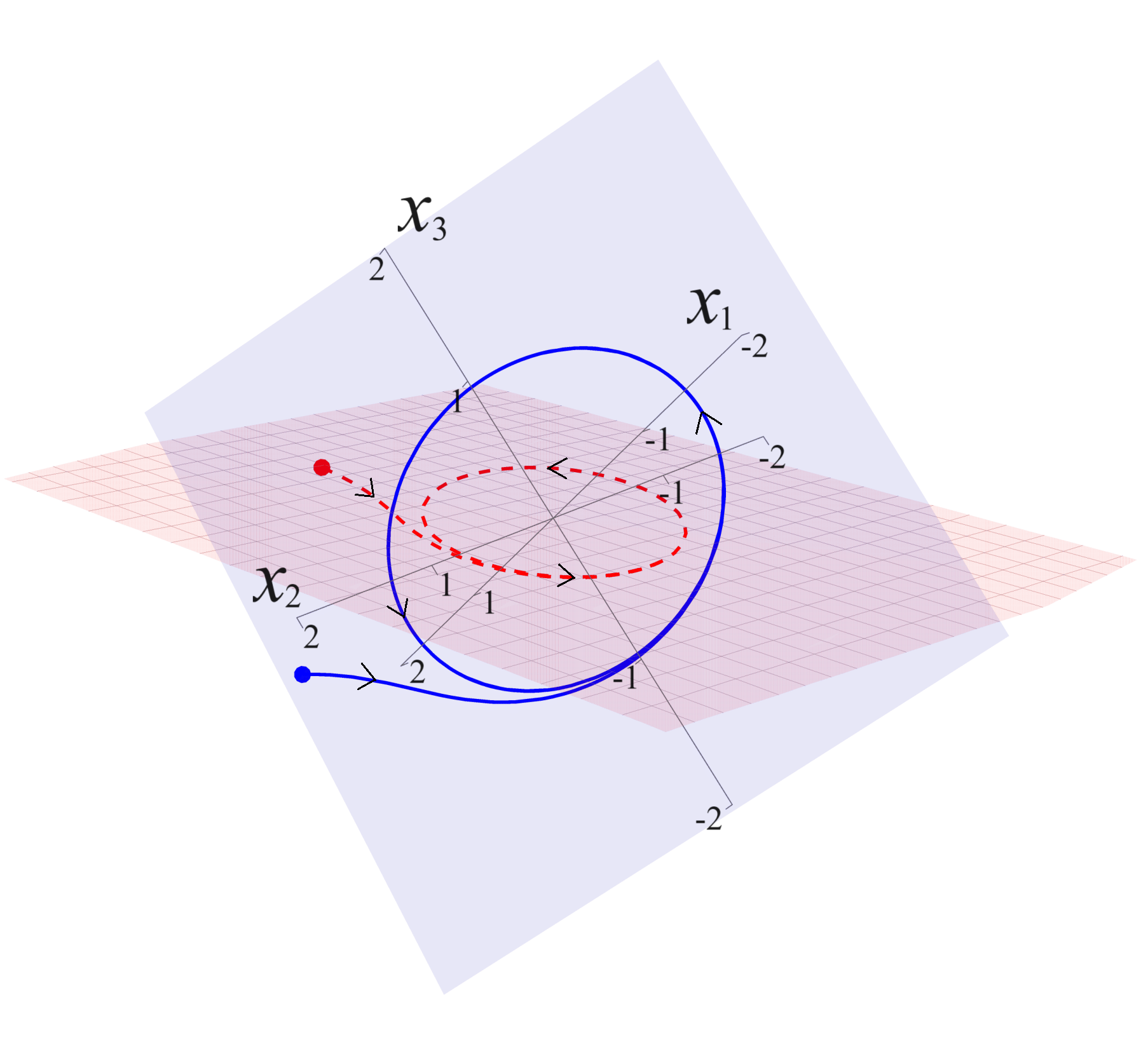}
     \caption{(Color online) The solid blue and dashed red lines are projections of two trajectories 
     for the four-dimensional system with quaternionic truncation as described in the text. 
     The parameter values are $(\nu_1, \nu_2, \nu_3) = (-1.6, 0.3, -0.2)$ for both orbits which have
     different initial conditions.}
      \label{fig:3}
\end{figure}
\end{center}
With these relations between the components the (symmetric) matrix $\mathbb{M}^2 
= - \nu^2 \mathbf{1}$, with $\nu^2 = \nu_1^2 + \nu_2^2 + \nu_3^2$. Consequently from Eq.\eqref{eq:bplane} one can see that 
$\boldsymbol{\dot{\elle}} = 0$ at all times  and the trajectory is thus constrained to remain on the initial 2-plane $\x_0\wedge\dot{\x}_0$. 
Since the coordinates $x_i$ are themselves the normal modes the asymptotic equations of motion are simply
$\Ddot{\xi}_i = -\nu^2 \xi_i$.
The solution
\begin{equation}
\xi_i(t) = \xi_i^{0} \cos{\nu t} + \frac{1}{\nu} \sum_{j\ne i} \mu_{ij}\xi_j^{0} \sin{\nu t},
\label{eq:15}
\end{equation}
defines a great circle at the intersection of the 2-plane $\mathsf{P}_2(0)$ with $\mathbb{S}^3$. 
This 2-plane is also the basin 
of attraction of the limit cycle. The projection of the trajectory to 
the three-dimensional subspace 
$(x_1, x_2, x_3)$ can be shown to lie on the plane $\sum_i c_i x_i= 0$, where 
\begin{equation*}
c_i = \!\!\sum_{\stackrel{j,k=1}{j\neq k,\text{\&}\,j,k \neq i}}^3\!\!\!\! \Big(\frac{1}{2} \nu_i \big( (\xi_j^0)^2 + (\xi_k^0)^2\big) - \nu_j (\xi_i^0\xi_j^0 -
\epsilon_{ijk}\xi_k^0\xi_4^0\big)\Big),
\end{equation*}
which is in fact the projection of the 2-plane $\x_0\wedge\dot{\x}_0$ (see\  Fig.~\ref{fig:3}). This differs from the generic 
three-dimensional case where the motion is not constrained to remain on a 
particular 2-plane since $\boldsymbol{\ell}$ is not 
constant except when the initial point lies on an equatorial plane.\\

The term $(\varrho - r^2)\x$ in Eq.\eqref{eq:2} may be replaced by a more general vector valued function $\F(\x)$, 
different choices for which can lead to other, possibly solvable, dynamical systems 
in $D$ dimensions. For instance, $\F(\x) = 0$ describes uncoupled higher dimensional Kuramoto oscillators \cite{Ott,Boccaletti2}, 
while $\F(\x) = - (\varrho-|\x|^2)\x$ generalises the subcritical Hopf bifurcation to 
higher dimensions: the origin becomes a stable fixed point as $\varrho$ crosses zero,
and unstable periodic/quasiperiodic orbits are created. Initial points on the 
hypersphere $r=\sqrt{\varrho}$ yield orbits on the hypersphere, however, starting 
away from it either ends at the origin, or in an unbounded orbit. With the choice 
$\F(\x)=(1-|\x|^2)\x$, the $D$-dimensional system under coupling has been studied by 
Zou {\em et al} \cite{Kumar}. Indeed, any set of dynamical equations that support separability
(cf.~Eq.~\eqref{eq:ComplexEOM}) will be integrable and hence solvable. The choice 
$\F(\x)=(\varrho_1 - |\x|^2)(|\x|^2 - \varrho_2)\x$ is a case in point for which the 
solution may be expressed in terms of elliptic integrals. A stable 2-sphere of radius $\sqrt{\varrho_1}$ and an unstable 
sphere of radius $\sqrt{\varrho_2}$ (see Supplementary Material \S{2}), 
are generated at the bifurcation in this case. Non-separable $\F(\x)$ can be considered within this 
framework and these will also be of interest to study. \\

To summarise, {we have extended the Stuart-Landau oscillator using the framework of Clifford algebra and studied the
resultant exactly solvable oscillator system with rotational symmetry in $D$-dimensions that models Hopf bifurcation.} 
At the bifurcation point a new kind of transition is observed, from a stable fixed point to a periodic or quasiperiodic state 
with {\em extreme multistability}.  We obtain the exact solution and give a complete 
description of the attractors as well as their basins. Our framework is flexible 
and provides a consistent method of generalising dynamical systems to arbitrary dimensions. 
This opens up many promising avenues of future research.  The higher dimensional Hopf 
bifurcation has importance in areas such as mathematical biology, ecology and medicine 
\cite{Bodnar,Forys,Freedman}. In addition, the dynamics of higher dimensional coupled 
systems with selectively broken symmetries is of great interest and is part of our ongoing
research \cite{Gogoi2}. The asymptotic behaviour of ensembles of such oscillators and the 
effect of their interactions on the symmetries dictates the dynamics of their 
collective states \cite{Nirmal,Awadhesh}. Exploration of possible quantum mechanical analogues \cite{Kenta,Tanmay} 
of this model will be another natural extension of our work.\\
\newline
\hspace{0.5mm}PBG acknowledges financial support from the University of Delhi 
through the University Research Fellowship.  RG and RR thank the Dept. of Chemistry, 
IIT-Delhi, for hospitality during the time this work was initiated. DG is supported in part by 
MATRICS (MTR/2020/481), SERB, Govt.\ of India. We thank Ulrike Feudel and Koushik Ray for discussions. 


\begin{thebibliography}{99}
\bibitem{fn1} The asymptotic dynamics for $\varrho < 0$ is a damped motion that tends to 
the origin, a fixed point, while for $\varrho>0$ it is oscillatory along the circle $|z|^2=
\varrho$ with an angular frequency $\omega$.
\bibitem{fn2} The nature of the dynamics in Eq.~\eqref{eq:0} 
changes drastically if $z$ is replaced by $\bar{z}$ on the RHS. The latter case is not 
symmetric under a rotation of the plane. In $D$ 
dimensions, it is possible to find several variations: for example, with the replacement 
$\x \to -\e_i\x\e_i$ on the RHS of Eq.~\eqref{eq:2}, 
only $x_i$ changes sign, thereby reversing the orientation of all hyperplanes involving $\e_i$.
\bibitem{fn3} {\color{black}All points on a given cone will lead to the same torus attractor and will 
   therefore correspond to the same set of $\lceil{D/2}\rceil$ conserved quantities, namely $N=\lfloor{D/2}\rfloor$  for
   even $D$ and $N=\lfloor{D/2}\rfloor +1$ for odd $D$.}
\end{thebibliography}
\end{document}